\documentclass[aps,prb,reprint,amsmath,amssymb,superscriptaddress]{revtex4-2}

\usepackage{graphicx}% Include figure files
\usepackage{dcolumn}% Align table columns on decimal point
\usepackage{bm}% bold math
\usepackage{float}\usepackage{amssymb}
\usepackage{xcolor}
\usepackage{tcolorbox}
\usepackage{hyperref}% add hypertext capabilities
\usepackage{braket}

\begin{document}
	
	%\preprint{}
	\title{Thermodynamics and entanglement entropy of the non-Hermitian SSH model}% Force line breaks with \\
	%\thanks{A footnote to the article title}%
	
	\author{D.F. Munoz-Arboleda}
	\email{d.f.munozarboleda@uu.nl}
	\affiliation{Institute for Theoretical Physics, Utrecht University, Princetonplein 5, 3584CC Utrecht, The Netherlands}%
	
	\author{R. Arouca}
	\affiliation{Department of Physics and Astronomy, Uppsala University, Uppsala, Sweden}
	
	\author{C. Morais Smith}
	\affiliation{Institute for Theoretical Physics, Utrecht University, Princetonplein 5, 3584CC Utrecht, The Netherlands}%
	
	\date{\today}% It is always \today, today,
	%  but any date may be explicitly specified
	
	\begin{abstract}
		Topological phase transitions are found in a variety of systems and were shown to be deeply related with a thermodynamic description through scaling relations. Here, we investigate the entanglement entropy, which is a quantity that captures the central charge of a critical model and the thermodynamics of the non-reciprocal Su-Schrieffer-Heeger (SSH) model. Although this model has been widely studied, the thermodynamic properties reveal interesting physics not explored so far. In order to analyze the boundary effects of the model, we use Hill's thermodynamics to split the grand potential in two contributions: the extensive one, related to the bulk, and the subdivision one, related to the boundaries. Then, we derive the thermodynamic entropy for both, the edges and the bulk and the heat capacity for the bulk at the topological phase transitions. The latter is related to the central charge when the underlying theory is a conformal field theory, whereas the first reveals the resilience of the topological edge states to finite temperatures. The phase transition between phases that are adiabatically connected with the Hermitian SSH model display the well-known behaviour of systems within the Dirac universality class, but the transition between phases with complex energies shows an unexpected critical behavior, which signals the emergence of an imaginary time crystal.
	\end{abstract}
	
	%\keywords{Suggested keywords}%Use showkeys class option if keyword
	%display desired
	\maketitle
	
	%\tableofcontents
	
	\section{Introduction}
	Topological states of matter offer a rich playground for exploring novel phenomena that could lead to advances in technological applications. Examples range from the quantum Hall effect in defining the standard of measurements of the quantum of conductance $e^2/h$ with unprecedented precision \cite{AndoMatsumotoUemura-JPSJ1975, Klitzing-PRL1980}, to the use of the quantum spin Hall effect in spintronics \cite{JUNGWIRTH-NM2012, KANEMELE-PRL2005, KANEMELE-PRL20052, QIXIAO-PT2010, HASANKANE-RMP2010}, to Majorana fermions as robust qubits in quantum computation \cite{ECMARINO-QFTCMP, PACHOS-ITQC}, to cite just a few. The theoretical foundation of topological phases of matter guided us to a generic classification of topological systems in terms of the tenfold way, in which the symmetries and dimension of a non-interacting model determine to which topological class they belong \cite{ATLAND-PRB1997,SHINSEI-JHEP2006}.
	
	In recent years, non-Hermitian systems have also garnered increasing interest due to their connections to fundamental concepts including topology and symmetry breaking. Non-Hermiticity arises naturally in non-equilibrium open systems \cite{Bender-RPP2007}, mechanical metamaterials \cite{Brendel-PRB2018}, and systems with dynamical instabilities \cite{KatsuraHosho-PRL2010, Kawaguchi-PR2012}. Contrarily to the Hermitian systems, the non-Hermitian ones exhibit features like non-orthogonal eigenstates \cite{Rotter-JPA2009, ASHIDAGONG-AP2020}, complex energy spectrum \cite{Bender-PRL1998}, point gaps, exceptional points \cite{BERRY-CJP2004,BERGHOLTZBUDICH-RMP2021}, non-Bloch band collapse (nBBC) \cite{MartinezAlvarez-PRB2018, YAOWANG-PRL2018}, and the non-Hermitian skin effect (nHSE) \cite{KUNSTEDVARDSSON-PRL2018, LEETHOMALE-PRB2019,KunstDwivedi2019,OKUMAKAWABATA-PRL2020}. In adition, non-Hermiticity changes the properties of topological boundary states, which enriched the topological classification from a 10-fold way \cite{Kitaev-AIPCP2009} to a 38-fold way \cite{KAWABATASHIOZAKI-PRX2019}. For example, non-unitary operations cannot be performed in Hermitian systems but are possible in non-Hermitian ones. Since the eigenstates of the system no longer form a complete set, a biorthogonal basis must be used, which is composed of a combination of left and right eigenstates \cite{MOISEYEV-NHQM,ASHIDAGONG-AP2020}.    
	%The study of the thermodynamics of small systems with finite-size effects have become of interest because of the perspective that can give to the topological phase transitions \cite{QUELLECOBANERA-PRB2016,KEMPKESQUELLE-SR2016,CATSQUELLE-PRB2018,AROUCAKEMPKES-PRR2020}. For this kind of systems, it is common to use the approach of Terell Hill (Hill thermodynamics)\cite{HILL-TSS}. 
	
	A paradigmatic model for investigating, theoretically and experimentally, non-Hermitian topological effects is the non-Hermitian Su-Schrieffer-Heeger (nH-SSH) model \cite{YaoWang2018, KUNSTEDVARDSSON-PRL2018,KunstDwivedi2019, GHATAKBRANDERBOURGER-PNAS2020, HelbigHofman2020, EekMoustaj2022, SlootmanCherifi2024}. Although the characterization of the Hermitian SSH model is done through winding numbers \cite{Ryu_2010, JKASBOTH-SCTI}, for a nH-SSH model with NHSE like the non-reciprocal SSH model, one cannot directly use momentum space to compute this invariant because the system has different spectra, depending on boundary conditions. For open boundary conditions (OBC), the topology of the nH-SSH model can be determined using the biorthogonal polarization \cite{KUNSTEDVARDSSON-PRL2018} or by computing the winding number calculated through the surrogate Hamiltonian \cite{KAWABATASHIOZAKI-PRX2019}. This characterization provides a phase diagram for the model, with four different phases: trivial and topological, in which the pseudo-Hermitian symmetry (PH-symmetry) is protected or broken \cite{KAWABATASHIOZAKI-PRX2019,AROUCALEE-PRB2020}.
	
	Recently, it was conjectured whether topological phase transitions could be identified with thermodynamic observables \cite{QUELLECOBANERA-PRB2016}. Since the system cannot be described using Landau theory because the order parameter is non-local, it has been long believed that this framework would be inadequate. Nevertheless, it was shown that it is possible to identify all the topological phase transitions in 1D, 2D, and 3D models by combining the Ehrenfest classification with Hill thermodynamics \cite{QUELLECOBANERA-PRB2016,KEMPKESQUELLE-SR2016,CATSQUELLE-PRB2018,AROUCAKEMPKES-PRR2020}. The Ehrenfest classification relates the order of the phase transition to the order of the derivative of the grand potential with respect to a parameter driving the phase transition. In recent years, it was shown that the critical exponents corresponding to the scaling of the system are related to the order of the phase transitions via the Josephson's hyperscaling relation \cite{Continentino-QSMBS,Continentino-PR1994}. This relation is in agreement with the Ehrenfest classification and can be extended also for the case of fractional-order phase transitions \cite{AROUCALEE-PRB2020}.
	
	In the Hill thermodynamics framework \cite{HILL-TSS}, the internal energy of a small system does not scale linearly with the volume of the whole system. To capture the non-linearity of the internal energy in terms of the volume, Hill considered an ensemble of equivalent, distinguishable, and independent subsystems. By investigating how the energy changes when more subsystems are added to the ensemble, it is possible to determine the most general form of the energy that accounts for the finite-size and boundary effects. In addition, there is a change in the energy of the ensemble when we vary the volume of each subsystem. These two mechanisms allow to modify the volume of the ensemble, and obtain all the observables for the bulk and boundary independently. The order of the phase transitions, the thermodynamic entropy, and the specific heat are some of the properties of interest to characterize the system. 
	
	Phase transitions can be identified not only through thermodynamics but with CFT \cite{Zamolodchikov-RMP1989}, statistical field theory \cite{Aizenman-JSP1987},  and entanglement entropy \cite{Chang-PRR2020}. Thus, an important tool for studying topological phases of matter is the entanglement spectrum \cite{LiHaldane-PRL2008}, which is derived from the reduced density matrix of a subregion of the system. Using the correlation function of the subregion \cite{Peschel-JPA2003}, it is possible to compute the entanglement spectrum and entanglement entropy for noninteracting topological systems \cite{Fidkowski-PRL2010,Turner-PRB2011}. In the case of Hermitian systems, the correlation function not only corresponds to topological edge states but also reveals additional ground state properties like the Zak phase, even in phases that are not topological \cite{Ortega-Taberner-PRB2021}. There is also a correspondence between the topological edge states with OBC and the $1/2$ modes of the correlation function. Extending this concept to non-Hermitian systems is challenging due to difficulties in defining the ground state and reduced density matrix. Recently, Herviou \textit{et al.} \cite{Herviou-PRA2019} generalized the correlation function to non-Hermitian systems. 
	
	In this work, we study the thermodynamics and entanglement entropy of the nH-SSH model. For pedagogical reasons, we first consider the thermodynamic entropy and heat capacity of the Hermitian SSH model using the Hill formalism. We find that at low temperatures, the Hermitian SSH model is in agreement with a CFT with central charge $c = 1$. Then, we investigate the nH-SSH model with OBC at zero temperature, and identify the trivial and topological phases through the entanglement entropy. A CFT was identified at the transition between the trivial and topological phases for both, the PH-symmetry protected and broken regimes. For the first, a central charge $c=1$ is found, whereas for the second, the entanglement entropy suggests that $c = 2$. Then, we investigated the thermal entropy and heat capacity in all phases. In the PH-symmetry protected phases, we find that the thermal entropy behaves similarly to the Hermitian SSH model. The heat capacity corroborates that the nH-SSH model is in agreement with a CFT with $c = 1$ at low temperatures. On the other hand, at the transition between the PH-symmetry broken phases, we find an oscillatory behavior of the heat capacity in terms of the inverse of the temperature, which can be associated with the emergence of an imaginary time crystal. Curiously, there are also oscillations in the T = 0 entanglement entropy as a function of the logarithm of the subsystem size, but those are damped in amplitude and frequency, and vanish in the thermodynamic limit, leaving a central charge $c=2$. Arguably, these oscillations have a different origin than the well-defined periodic peaks observed in the heat capacity at low temperatures.         
	
	This paper starts in section~\ref{HTs} with a brief introduction about the Hill thermodynamics formalism, which will be the theoretical framework for the models in study. In section~\ref{SSH}, the Hermitian SSH model is introduced and, its thermodynamic properties, such as thermal entropy and specific heat are calculated. It is then shown how they can be related to topological phases at finite temperature. In section~\ref{NRNHSSH}, we present our main results. First, we introduce the non-Hermitian non-reciprocal SSH model, discuss the different phase transitions that arise in this model at zero temperature, and compute the entanglement entropy. In particular, we focus on the order of the phase transition between PH-symmetry protected (broken) trivial (topological) and topological (trivial) phases. Then, the results for the thermal entropy and specific heat are presented. We show how the thermal entropy of the edges can reveal the resilience of the topological edge states to finite temperature, and how the specific heat can be related to the central charge of a CFT theory for some of the critical points. The emergence of an imaginary time crystal is shown at the transition from the PH-symmetry broken topological to trivial phase. Finally, in section~\ref{conclusions} we present the conclusions of our work.
	
	\section{Hill Thermodynamics}\label{HTs}
	In the usual thermodynamics, the internal energy of a system is written as $E_t=TS_t -pV_t+\mu N_t$, where $E_t$ is the total internal energy, $S_t$ is the total entropy, $V_t$ is the total volume and $N_t$ is the total number of particles in the whole ensemble. Terell Hill was interested in studying the thermodynamics of small systems. He considered an ensemble of $\mathcal{N}$ equivalent, distinguishable and independent subsystems, all characterized by the volume $V$, temperature $T$ and chemical potential $\mu$ \cite{HILL-TSS}. In the thermodynamic limit $\mathcal{N}\rightarrow \infty$, the whole ensemble is a macroscopic thermodynamic system, although each individual subsystem is not. Then, the total energy of the system will also depend on $\mathcal{N}$ as
	\begin{equation}
		dE_t=TdS_t -pdV\mathcal{N}+\mu dN_t-\hat{p}Vd\mathcal{N},\label{2-1}
	\end{equation}
	where $p$ is the mean pressure, $\hat{p}$ is the pressure per subsystem, and the term $-\hat{p}Vd\mathcal{N}$ is related to the work done by the subsystems, i.e. is a parameter that shows how the total energy changes when the number of subsystems $\mathcal{N}$ changes. Increasing the number of subsystems with volume $V$ will also increase the volume of the whole ensemble. For macroscopic systems, $p=\hat{p}$, but for small systems they will differ. In the grand canonical ensemble, $\mu$, $T$, and $V$ are constant, and we can integrate~(\ref{2-1}) to obtain the total energy
	\begin{equation}
		E_t=TS_t +\mu N_t-\hat{p}V \mathcal{N}. \label{2-2}
	\end{equation}
	Let us evaluate the average of the total energy, entropy, and number of particles by dividing Eq.~(\ref{2-2}) by $\mathcal{N}$ and taking into account that $E_t=\bar E \mathcal{N}$, $S_t= S \mathcal{N}$ and $N_t=\bar N \mathcal{N}$. Then
	\begin{equation}
		\displaystyle  \bar{E}=TS +\mu \bar{N} -\hat{p}V. \label{2-3}
	\end{equation}
	Equation~(\ref{2-3}) has the same functional form as for macroscopic systems; the only difference is that for the usual thermodynamics there is $p$ and for the Hill description, $\hat{p}$. This  small difference seems to be irrelevant, but actually has an important consequence, since $\hat{p}$ may depend on $V$. Then, the energy $\bar{E}$ is not extensive in $S$, $V$ and $\bar{N}$, as in the case of the macroscopic systems.  Now,  Eq.~(\ref{2-3}) may be rewritten as
	%An important feature of Hill thermodynamics is that all the thermodynamic identities remain intact. 
	%Now, inserting Eq.(\ref{2-3}) in its differential form and dividing all terms by $\bar{N}$, we obtain $d\bar{E}=TdS+\mu d\bar{N}-\hat{p}dV$. Integrating this expression in the grand canonical ensemble leads to the internal energy per subsystem
	\begin{equation}
		\bar{E}=TS + \mu \bar{N}- pV + \mathcal{X}, \label{2-4}
	\end{equation}
	where $\mathcal{X} =(p-\hat{p})V$ denotes the subdivision potential. Equation~(\ref{2-4}) captures the full non-extensive behavior of a finite-size system in the grand canonical ensemble.
	
	The Hill formalism can also be used to describe topological models, where the subdivision potential provides a thermodynamic description of the edge states \cite{QUELLECOBANERA-PRB2016}. Let us consider a system of finite size $V$ in contact with an environment at temperature $T$ and chemical potential $\mu$. Then, the grand potential is
	\begin{eqnarray}
		\Omega &=& \bar{E} -TS -\mu \bar{N} \nonumber \\
		&=& \Omega_{ext} + \Omega_{n-ext}, \label{2-5}
	\end{eqnarray}
	where $\Omega_{ext}=-pV$ is the extensive part of the grand potential and $\Omega_{n-ext}=\mathcal{X}$ is the non-extensive or the subdivision potential. It is useful to compute the grand partition function to obtain the thermodynamic properties
	\begin{eqnarray}
		\displaystyle \mathcal{Z}&=&\sum_{q}\left <\psi_q\right |\rm{e}^{\beta(\hat{H}-\mu\hat{N})}\left |\psi_q\right > \\ \nonumber
		&=&\textrm{Tr} \left\{\rm{e}^{-\beta (\hat{H}-\mu\hat{N})}\right\}, \label{2-6}
	\end{eqnarray}
	where $\beta=1/(k_BT)$, with $k_B$ the Boltzmann constant, and $\hat{N}=a^{\dagger}a$ is the number operator. We will focus on free fermionic systems with single particle spectrum $\mathcal{E}_i$. Taking into account that a free fermionic system has two occupation numbers (0,1), we obtain the partition function
	\begin{eqnarray}
		\displaystyle \mathcal{Z}
		=\prod_{i}\left(1+\rm{e}^{-\beta(\mathcal{E}_i-\mu)}\right). \label{2-7}
	\end{eqnarray}
	The grand potential, defined as $\Omega=-\beta^{-1}\ln \mathcal{Z}$, can then be promptly computed using Eq.~(\ref{2-7}),
	\begin{equation}
		\displaystyle \Omega(T, \mu) =-k_BT\sum_{i}\ln \left[1+\rm{e}^{-(\mathcal{E}_i-\mu)/k_BT}\right]. \label{2-8}
	\end{equation}
	
	By calculating the grand potential for different system sizes, we are able to separate the bulk (extensive) and boundary (non-extensive) contributions. Then, by taking their derivatives, we can obtain physical observables, such as the heat capacity or the density of states. Next, we introduce the Hamiltonian of the model that we are going to investigate. For the sake of simplicity, we first analyse the Hermitian SSH model before considering its non-Hermitian counterpart.
	
	\section{Su-Schrieffer-Heeger (SSH) model}\label{SSH}
	\subsection{The model}\label{MSSH}
	
\begin{figure}[tbh]
		\centering
		\includegraphics[scale=0.3]{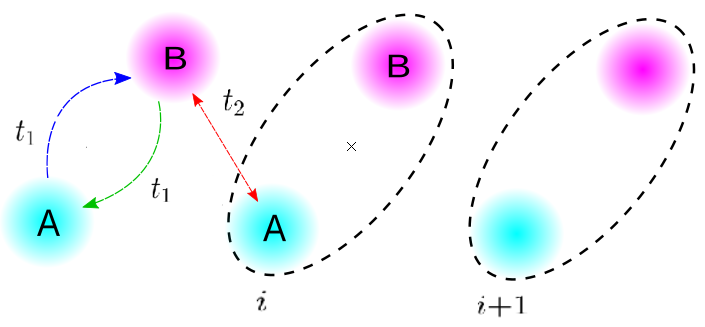}
		%\captionsetup{justification=raggedright}
		\caption{Sketch of the Hermitian SSH model. The dashed lines indicate every cell in the lattice. The blue and green solid lines describe the intracell hopping parameter $t_1$. The red solid line represents the intercell hopping parameter $t_2$. }
		\label{SSH_Model}
	\end{figure}

	The SSH model (see sketch in Fig.~\ref{SSH_Model}) is a paradigmatic one-dimensional model which presents a topological phase transition \cite{SuSchriefferHeeger-PRL1979,ECMARINO-QFTCMP}.
	It consists of a 1D chain with two sublattices, A(cyan) and B(magenta). There are two hopping parameters, $t_1$ and $t_2$, which describe, respectively, the intracell and the intercell hopping amplitude. The Hamiltonian of the model reads
	\begin{equation}
		H=\displaystyle \sum_{i=1}^{N}\left(t_1 a_i^{\dagger}b_i+t_2a_{i+1}^{\dagger}b_i+\textnormal{h.c.}\right), \label{3-1}
	\end{equation}  
	where $a_{i}^{\dagger}$ and $b_{i}^{\dagger}$ ($a_{i}$ and $b_{i}$) create (destroy) a fermion on the A and B site of the $i$ unit cell. For periodic boundary conditions (PBC), the Bloch Hamiltonian corresponding to Eq.~(\ref{3-1}) is given by
	\begin{equation}
		H_{\textnormal{bulk}}(k)=\begin{pmatrix}
			0&t_1+t_2\rm{e}^{-ik}\\
			t_1+t_2\rm{e}^{ik}& 0
		\end{pmatrix}, \label{3-2}
	\end{equation}
	and the energy reads
	\begin{equation}
		E(k)=\pm \sqrt{t_1^2+t_2^2+2t_1t_2\cos k}.
	\end{equation}
	
	The Bloch Hamiltonian for any model with two internal states per unit cell can be described as a vector $\bf{d}$ multiplied by the Pauli matrix vector $\mathbf{\sigma}$
	\begin{equation}
		H(k)=d_{x}(k)\hat{\sigma}_{x}+d_{y}(k)\hat{\sigma}_{y}=d_{0}(k)\hat{\sigma}_{0}+\mathbf{d}(k)\cdot \mathbf{\sigma} \label{3-3},
	\end{equation}
	where $d_0(k)=d_z(k)=0$, $d_x(k)=t_1+t_2\cos k$ and $d_y(k)=t_2\sin k$.
	Solving the Schrödinger equation $\displaystyle H(k)\left|\psi_n(k)\right>=E_n(k)\left|\psi_n(k)\right>$ for the Bloch Hamiltonian, we can have a representation of the states $\displaystyle \left|\psi_{n}(k)\right>$ in terms of $\bf{d}$$(k)=(d_x(k),d_y(k),d_z(k))$
	\begin{equation}
		\displaystyle \left|\psi_{\pm}(k)\right>= \begin{pmatrix}
			\pm \frac{d_x(k)-id_y(k)}{|\mathbf{d}(k)|}\\
			1
		\end{pmatrix} \label{3-4}.
	\end{equation}
	
	There is an energy gap $\Delta=\textnormal{min}_{k}E(k)=|t_1/t_2|-1$ which separates the last filled band from the first empty one. When $t_1/t_2=1$, the gap closes, and there is a topological phase transition. The topological phase can be characterized through topological invariants calculated from bulk modes of the system. In 1D, symmetry protected topological modes appear usually at zero energy. The above is a manifestation of the bulk-boundary correspondence  between the presence of topological modes at the boundary of the system with a topological invariant in the bulk. These topological phases can be identified by the associated Berry phases and the winding number ($W$) \cite{JKASBOTH-SCTI, KEMPKES-2016, CATS-2017}. The last relates how many times the vector $\mathbf{d}(k)$ winds around the origin in the $d_x$ and $d_y$ plane. The winding number can only have integer values,
	\begin{equation}
		\displaystyle W = \frac{1}{2\pi}\int_{-\pi}^{\pi}i\left<\psi(k)\right|\partial_{k}\left|\psi(k)\right> dk\label{3-5}.
	\end{equation}
	
	Equation~(\ref{3-5}) reveals some interesting properties for the winding number: if $W=0$, the vector $\mathbf{d}(k)$ does not wind the origin; if $W=1$, the vector $\mathbf{d}(k)$ winds around the origin once in a counterclockwise manner. A different value for $W$ indicates that $\mathbf{d}(k)$ touches the origin. Then, the winding number can capture the topology of the system. When the gap closes, the winding number $W$ becomes ill-defined. If $t_1/t_2<1$ ($t_1/t_2>1$), the SSH model hosts a gap which is related with a topological phase (trivial phase). The presence of the topological (trivial) phase can also be inferred from the winding number $W=1$ ($W=0$).
	
	\subsection{Thermal entropy and heat capacity}\label{SSHES}
	\begin{figure}[tbh]
		\centering
		\includegraphics[scale=0.6]{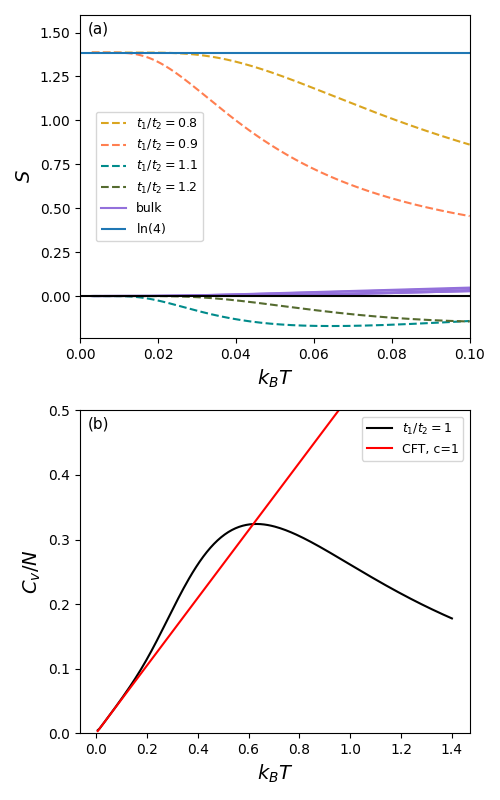}
		%	\captionsetup{justification=raggedright}
		\caption{Entropy and specific heat of the SSH model. (a) Thermal Entropy of the SSH model. Purple lines correspond to the bulk, color dashed lines correspond to the edge for different values of $t_1/t_2=\{0.8,0.9,1.1,1.2,1.3\}$, the blue line corresponds to $\ln(4)$. (b) Specific heat of the SSH model (black line) at the phase transition $t_1/t_2=1$. The red line shows a linear behavior of the specific heat for low-$T$ temperatures, with a slope $C_v/Nk_B T \approx \pi/3$, which means a central charge of $c=1$. These plots were created with a system size $L=200$.}
		\label{EntropyCv_SSH}
	\end{figure}
	It was originally shown in Refs.~\cite{KEMPKESQUELLE-SR2016, KEMPKES-2016} that the phase transitions for the SSH model can be described using the Hill thermodynamics formalism. %In Fig.(\ref{SSH_THERMDWDT}) we can see both, the gap closing and the appearance of the zero modes. 
	Taking into account the spectrum of the model for PBC and OBC, one can obtain the grand canonical potential [see Eq.~(\ref{2-6})]. Using Hill thermodynamics, it is possible to separate the different contributions for the bulk and the boundary. The extensive contribution is obtained through the PBC system $\omega_{ext}=\Omega_{PBC}/(2L)$, where $L$ is the number of unit cells, while the non-extensive contribution of the OBC system is $\omega_{n-ext}=\Omega_{OBC}-\Omega_{PBC}$.
	The entropy $S=-\partial \Omega_i/\partial T$ must be calculated for the bulk and edge separately. Using Eq.~(\ref{2-8}) and $\mu=0$, we can compute both contributions
	\begin{equation}
		\displaystyle S=\sum_{i} k_B\ln \left(1+\rm{e}^{\mathcal{E}_i/k_BT}\right)-\frac{1}{T}\mathcal{E}_i\left(\frac{\rm{e}^{\mathcal{E}_i/k_BT}}{1+\rm{e}^{\mathcal{E}_i/k_BT}}\right), \label{3-6}
	\end{equation}
	where $\mathcal{E}_i$, denotes the energy of the system. 
	
	Figure~\ref{EntropyCv_SSH}(a) shows the entropy of the SSH model for different values of the parameter $t_1$ near the phase transition. The purple lines depict the bulk contribution to the entropy for different values of $t_1/t_2$. They start at zero and grow linearly, as expected. The color dashed lines represent the entropy of the edge. An interesting behavior may be observed near the phase transition at $t_1/t_2=1$: for $t_1/t_2<1$, the edge contribution goes from the initial value $\ln(4)$ in the topological phase for $T=0$ to zero as the temperature is increased. The closer the value of $t_1/t_2$ is to the critical point $t_1/t_2=1$, the earlier the entropy decays from $\ln(4)$. This $\ln(4)$ value arises because in the SSH model there are 2 edge dimerized modes, unlike the Kitaev chain, which has 2 Majorana edge modes and therefore starts at $\ln(2)$ \cite{KEMPKESQUELLE-SR2016}. For the trivial phase, $t_1/t_2>1$, the edge entropy starts at zero and becomes negative upon increasing the temperature. One observes that the closer the parameter $t_1/t_2$ is to the critical point, the earlier the entropy decays and becomes negative. This means that the edges act as a refrigerator for the system \cite{YUNTFADAIE-PRB2020}. The total entropy remains positive because the bulk entropy, which is positive, dominates.  
	
	We calculate also the heat capacity $C_V=T\partial S/\partial T$. Deriving Eq.(\ref{3-6}) with respect to $T$, we obtain 
	\begin{equation}
		C_V=\frac{1}{T^2}\sum_{i}\mathcal{E}_{i}^2\frac{\rm{e}^{\mathcal{E}_i/k_BT}}{\left(1+\rm{e}^{\mathcal{E}_i/k_BT}\right)^2}. \label{3-7}
	\end{equation}
	
	For 1D quantum systems, $C_V$ is proportional to the central charge $c$ as $C_V=\pi c k_B^2TN/(3\hbar v)$ with $v$ the velocity of the excitations and $N$ the length of the system (here we relate the length with the amount of sites in the lattice and set $k_B/(\hbar v)=1$). The bulk specific heat $C_v/N$ for $t_1/t_2=1$ is plotted in Fig.~\ref{EntropyCv_SSH}(b) with a black solid line. At low temperatures, the specific heat shows a linear behavior in $T$, $C_V/N=\pi k_B T/3$, see the red line in Fig.~\ref{EntropyCv_SSH}(b). The slope of the red line indicates that the central charge $c=1$, as predicted by the CFT of the system \cite{Pasquale-JSM2004}. Our results show the behavior of the heat capacity not only at zero temperature but also at finite temperatures. 
	
	\section{Non-reciprocal Non-Hermitian SSH model}\label{NRNHSSH}
	\subsection{The model and its phase diagram}\label{MPHD}
	
	Now, we will study the non-reciprocal nH-SSH model. The non-Hermiticity of the model will have some important consequences. One of those is that the eigenstates of the Hamiltonian no longer form a complete set. Thus, we need to use a biorthogonal basis, composed of the eigenstates $\displaystyle \left | \psi^L\right >$ and $\displaystyle \left | \psi^R\right >$, where $L$ ($R$) denotes left (right) \cite{MOISEYEV-NHQM, ASHIDAGONG-AP2020}. In general, for this set of eigenstates, we obtain
	\begin{eqnarray}
		\displaystyle H\left | \psi_n^R\right >&=&E_n\left | \psi_n^R\right >\\ \nonumber
		H^{\dagger}\left | \psi_n^L\right >&=&E_n^*\left | \psi_n^L\right >.
	\end{eqnarray}
	
	Here, the biorthogonal condition is imposed
	\begin{equation}
		\displaystyle \braket{\psi_n^L|\psi_m^R} = \delta_{nm}.
	\end{equation}
	
	Breaking the Hermiticity introduces also complex eigenenergies, which lead to novel effects. One of these is the accumulation of modes at the border of the system, the so-called nHSE \cite{ZHANG-CP2021,LEETHOMALE-PRB2019,SONGYAOWANG-PRL2019,OKUMAKAWABATA-PRL2020,LILEEMU-PRB2019}. These phenomena do not take place for closed systems, but arise in an effective description of open systems \cite{BREUERPETRUCCIONE-TOQS, BERGHOLTZBUDICH-RMP2021, PhysRevLett.127.070402}.
	
	The non-reciprocal nH-SSH model differs from the SSH model due to the intracell hopping parameters, which are different if the electrons hop from A to B, ($t_1-\delta$), blue dashed line in Fig.~\ref{SSH_Model} or from B to A, ($t_1+\delta$), green dashed line in Fig.~\ref{SSH_Model}. Notice that if $\delta=0$, the usual Hermitian SSH model is recovered. Thus, $\delta$ is a parameter that controls the non Hermiticity of the system \cite{KUNSTEDVARDSSON-PRL2018,YAOWANG-PRL2018,YINJIANG-PRA2018,GHATAKBRANDERBOURGER-PNAS2020,ADLAKHAMOGHADDASZADEH-SR2020}. Let us write $t_+=t_1+\delta$ and $t_-=t_1-\delta$. Then, the Hamiltonian of the non-reciprocal nH-SSH model reads
	\begin{equation}
		H=\displaystyle \sum_{i=1}^{N} t_+a_i^{\dagger}b_i+ t_-b_i^{\dagger}a_i+t_2\left(a_{i+1}^{\dagger}b_i+\textnormal{h.c.}\right). \label{3-8}
	\end{equation} 
	
	\begin{figure}[tbh]
		\centering
		\includegraphics[scale=0.6]{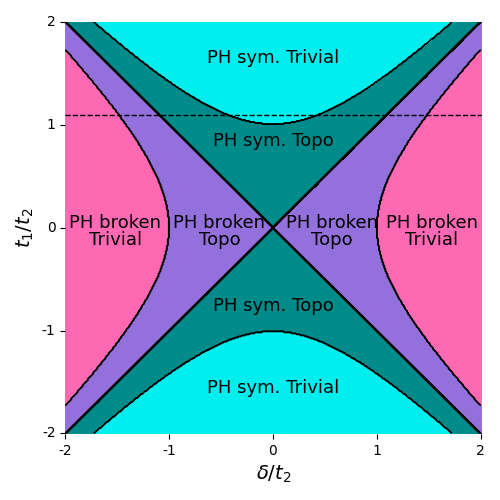}
		%	\captionsetup{justification=raggedright}
		\caption{Phase diagram of the nHSSH model with OBC. For $t_1=1.1$ (dashed black line), all phases may be obtained upon tuning $\delta/t_2$.}
		\label{NHSSH_OBC_PD}
	\end{figure}
	
	\begin{figure*}[tbh]
		\centering
		\includegraphics[scale=0.5]{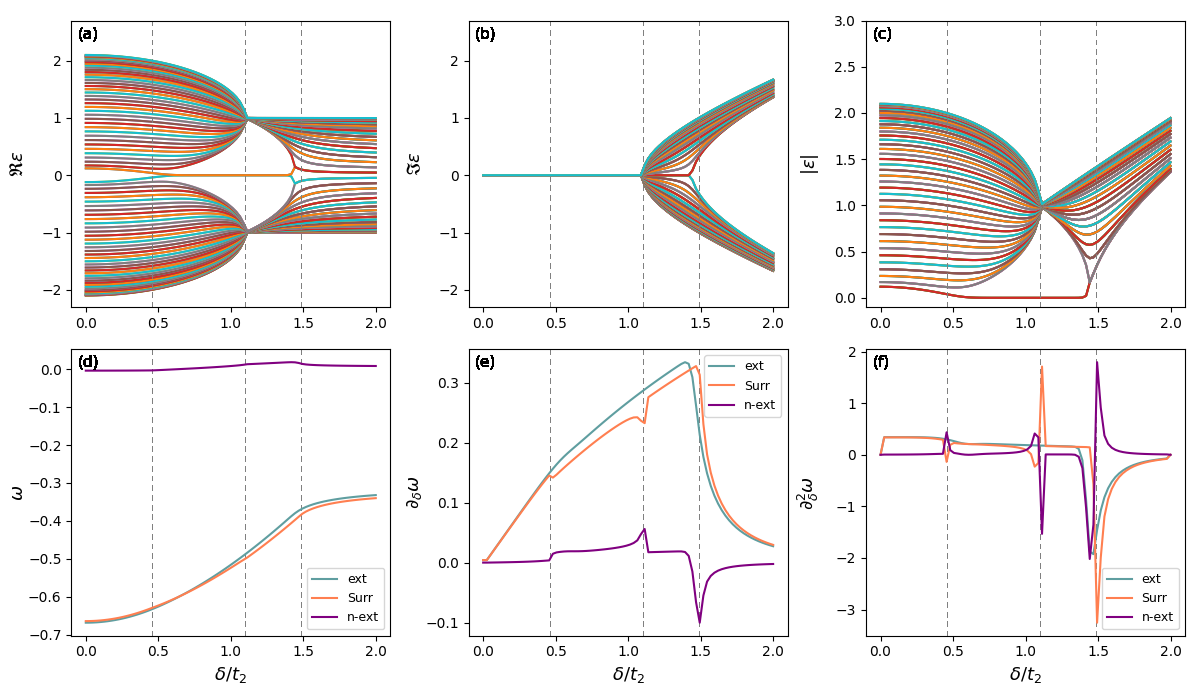}
		%\captionsetup{singlelinecheck=true, position=below, justification=raggedright}
		\caption{Comparison between the spectrum and grand potential of the non-Hermitian SSH model with OBC and the surrogate Hamiltonian for $t_1/t_2=1.1$. From (a)-(c), the spectra overlap, except for the absence of the zero-energy mode contribution for the surrogate Hamiltonian. (a) Real, (b) imaginary, and (c) absolute value of the energy. From (d)-(f), we show the surrogate (orange), extensive (cyan), and non-extensive (purple) contributions of (d) the grand potential, (e) its first derivative, and (f) its second derivative with respect to $\delta$. Dashed lines are drawn at the critical values of $\delta=0.46$, $\delta=1.1$, and $\delta=1.49$ to facilitate the visualization of the kinks. All calculations were made with $\mu=0$ and a system size $L=80$.}
		\label{NHSSH_Energyobcsurr_delta}
	\end{figure*}
	
	To overcome the nHSE, we need to change the basis into a new one, which is dependent on the position. This change gauges away the accumulation of modes in the edges, but preserves the spectrum of the OBC system. To this aim, one must perform a deformation of the momentum, $k\rightarrow k + i \kappa$. This deformation for the nH-SSH model is
	\begin{equation}
		\displaystyle \kappa=-\ln\sqrt{\frac{t_-}{t_+}},
	\end{equation}\label{3-9}
which generates a surrogate Hamiltonian $h_{\textnormal{surr}}(k)$ \cite{LEELI-PRB2020,HELBIG-NP2020}, where we fixed the hopping parameter $t_2=1$,
	\begin{equation}
		h_{\textnormal{surr}}(k)=\begin{pmatrix}
			0&t_- + \sqrt{\frac{t_-}{t_+}}e^{i k}\\
			t_+ + \sqrt{\frac{t_+}{t_-}}e^{-i k}& 0 
		\end{pmatrix}.\label{3-10}
	\end{equation}
	
	In terms of the original hopping parameters $t_1+\delta$ and $t_1-\delta$, the corresponding energy spectrum is given by
	\begin{equation}
		E_{\textnormal{surr}\pm}(k)=\pm\sqrt{1+t_1^2-\delta^2+2\sqrt{t_1^2-\delta^2}\cos k}.
		\label{eq_E_surr}
	\end{equation} 
	This spectrum is obtained from a periodic Hamiltonian, but reproduces exactly the spectrum of an OBC system, except for the zero modes that are located at the edges \cite{AROUCA-2021}.
	
	As in the Hermitian case, the topology for the nH-SSH model with OBC, described by $h_{\textnormal{surr}}$, can be captured by the winding number \cite{KAWABATASHIOZAKI-PRX2019}
	\begin{equation}
		W_{\textnormal{surr}} = \oint \frac{d k}{4i\pi} \mathbf{Tr}\left[\sigma_{z}h^{-1}_{surr}(k)\frac{dh_{surr}(k)}{dk}\right]. \label{3-11}
	\end{equation}
	
	Using the winding number from Eq.~(\ref{3-11}), we reproduce the phase diagram for OBC, obtained originally in Ref.~\cite{KAWABATASHIOZAKI-PRX2019} and extended in Ref.~\cite{AROUCALEE-PRB2020}. Here, we replot it because we rename some of the phases, see Fig.~\ref{NHSSH_OBC_PD}. Four different phases can be distinguished: The PH-symmetry protected trivial phases (cyan, $W_{surr}=0$) and the PH-symmetry protected topological phases (dark green, $W_{\textnormal{surr}}=1$) may be observed for the values $|\delta/t_2|< |t_1/t_2|$, whereas the PH-symmetry broken topological phases (purple, $W_{\textnormal{surr}}=1$) and the PH-symmetry broken trivial phases (pink, $W_{\textnormal{surr}}=0$) occur for $|\delta/t_2|> |t_1/t_2|$. 
	\subsection{Order of the phase transitions}\label{NHSSHTH}
	In the OBC case, there are no bulk localized modes, only skin modes at the edges of the system, which implies that the bulk-boundary correspondence is broken via the nHSE. In Figs.~\ref{NHSSH_Energyobcsurr_delta}(a)-(c), we show the real, imaginary, and absolute values of the energy for OBC and for the surrogate Hamiltonian. The graphs superpose; the only difference is that the zero energy lines are not visible for the surrogate Hamiltonian. As one varies $\delta$, an unconventional critical thermodynamic behavior with no Hermitian counterpart arises \cite{AROUCALEE-PRB2020}. Indeed, the asymmetric bunching of bands at the nBBC point $\delta_c/t_2=t_1/t_2=1.1$ \cite{LEELI-PRB2020} is visible in all types of spectra (real, imaginary and absolute). This is similar to a non-Hermitian flat band in a complex energy space, as all bands can intermix with divergent density of states. There is no gap closing at the nBBC, but it signals a transition nevertheless, leading to band metric discontinuities linked to the nHSE \cite{LEELI-PRB2020}. We see also the appearance and disappearance of zero modes between the PH-symmetry protected trivial and topological phase at $\delta_c/t_2=\sqrt{t_1^2/t_2^2-1}\approx0.46$ and between the PH-symmetry broken topological and trivial phase at $\delta_c/t_2=\sqrt{t_1^2/t_2^2+1}\approx1.49$, respectively. 
	Using Eq.(\ref{2-8}) at zero temperature, we compute the extensive ($\omega_{\textnormal{ext}}$) and non-extensive ($\omega_{\textnormal{n-ext}}$) contributions of the grand potential and its derivatives for OBC, as well as the surrogate ($\omega_{\textnormal{surr}}$) with respect to $\delta$, for $t_1/t_2=1.1$, see Figs.~\ref{NHSSH_Energyobcsurr_delta}(d)-(f). These values were chosen in order to access all the possible phase transitions shown in Fig.~\ref{NHSSH_OBC_PD} (see the dashed line). Without loss of generality, we will now concentrate on values of $\delta/t_2 > 0 $ and $t_1/t_2 > 0$. The other regions of the phase diagram are promptly accessed by symmetry. 
	The Ehrenfest classification associates the order of the phase transition to the order of the derivative of the grand potential that diverges or shows a discontinuity \cite{JAEGER-AHES1998}. Figures~\ref{NHSSH_Energyobcsurr_delta}(e) and (f) show the first and second derivatives of the grand potential for the $\omega_{\textnormal{ext}}$ (cyan), $\omega_{\textnormal{surr}}$ (orange) and $\omega_{\textnormal{n-ext}}$ (purple) contribution with respect to $\delta$. At the critical points $\delta_c/t_2\approx0.46$, $\delta_c/t_2\approx1.1$, and $\delta_c/t_2\approx1.49$, there seem to be kinks or discontinuities in the first and second derivatives, but it is difficult to determine the precise order of the phase transitions by applying just the Ehrenfest classification. Therefore, the Josephson's hyperscaling relation was used in Ref.~\cite{AROUCALEE-PRB2020}. It was found that for the critical point $\delta_c/t_2=0.46$, the phase transition has an integer order of 2, and for $\delta_c/t_2=1.49$, the phase transition has a fractional order of 1.5. On the other hand, for the nBBC point $\delta_c/t_2=t_1$, the order of the phase transition does not depend on the dimension of the system, such that the Josephson's hyperscaling relation is not satisfied \cite{AROUCALEE-PRB2020}.
	\begin{figure}[tbh]
		\centering
		\includegraphics[scale=0.6]{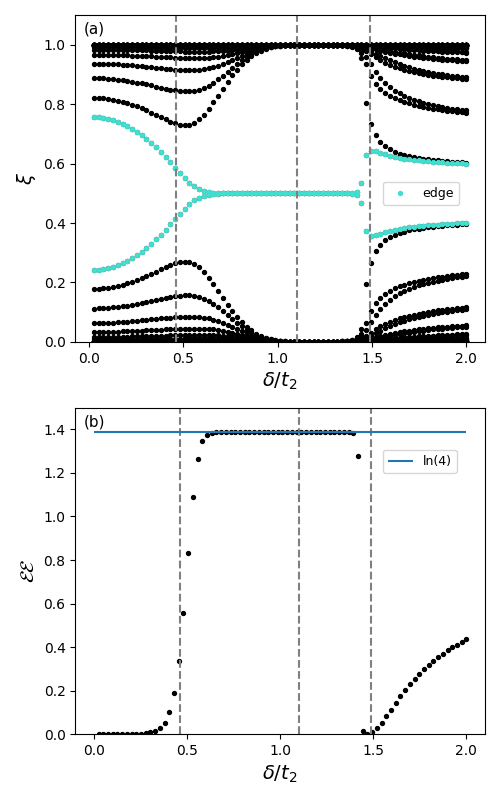}
		%	\captionsetup{justification=raggedright}
		\caption{(a) Correlation spectrum of the nH-SSH model. The cyan line denotes the edge states. (b) Entanglement entropy of the nHSSH model. The blue line corresponds to $\ln(4)$. Results were obtained for a system size $L=120$.}
		\label{Corr_Spec_NHSSH}
	\end{figure}
	\subsection{Entanglement entropy}
	In this section, we investigate the entanglement entropy of the non-reciprocal nH-SSH model at zero temperature. Previous studies of the entanglement entropy and entanglement Hamiltonian for the gain-loss nH-SSH model were done in Refs.~\cite{Chang-PRR2020,Pasquale-PRB2023,Pasquale-JST2024}. Here, we use a generalized definition of the reduced density matrix in the biorthogonal basis developed in Refs.~\cite{Brody-JPA2014, Weigert-PRA2003}. In particular, for non-interacting systems the entanglement entropy can be computed using the correlation functions of the biorthogonal basis \cite{Chang-PRR2020}. These correlation functions can be defined in different manners, using only left wavefunctions, right wavefunctions, or a combination of them \cite{Ortega-PRB2022}. In our case, we will use the latter   
	\begin{equation}
		\displaystyle C_{ij}^{LR}=\left< G^L \right| \phi_i^{\dagger}\phi_j \left| G^R \right>, \label{3-12}
	\end{equation}
	where $\left |G^L \right>$ ($\left |G^R \right>$) are the many-body left (right) eigenstates of the system and $\{\phi_i^{\dagger},\phi_j\}=\delta_{ij}$ are fermionic operators. For a subsystem $\mathcal{A}$, the entanglement entropy can be written as
	\begin{equation}
		\displaystyle \mathcal{EE} =-\sum_{\alpha}\left [ \xi_{\alpha}\ln \xi_{\alpha}+\left(1-\xi_{\alpha}\right)\ln\left(1-\xi_{\alpha}\right)\right], \label{3-13}
	\end{equation}
	where $\xi_{\alpha}$ are the eigenvalues of $C^{LR}$ \cite{Chang-PRR2020}. 
	Using Eqs.~\ref{3-12}, \ref{3-13}, and a similar numerical technique to the one developed in Ref.~\cite{Ortega-PRB2022}, we compute the correlation spectrum and entanglement entropy for the nH-SSH model with OBC. We consider a system with 120 sites. In Fig.~\ref{Corr_Spec_NHSSH}(a), we show the real part of the correlation spectrum with respect to the $\delta$ parameter for $t_1/t_2=1.1$. The edge (bulk) states are represented in cyan (black). For $\delta/t_2<0.46$, the edge states are gaped. In this region, the system hosts a PH-symmetry protected trivial phase. Between $\delta/t_2\sim0.5$ \footnote{Upon increasing the system size, the critical point moves to the left until it reaches the actual critical point at $\delta/t_2=0.46$. We show the plot for 120 sites in order to avoid noise} and $\delta/t_2=1.49$, the gap closes. In this regime, the eigenvalue of the correlation function acquires the value $\xi=0.5$, signalling a topological phase, see Fig.~\ref{NHSSH_Energyobcsurr_delta}. This result is in agreement with Refs.~\cite{Ortega-Taberner-PRB2021,Ortega-PRB2022}, which found a PH-symmetry protected and a PH-symmetry broken topological phases in the same regime of parameters. For $\delta/t_2>1.49$, we observe the reappearance of a gap, corresponding to a trivial phase. In Fig.~\ref{Corr_Spec_NHSSH}(b), we show the entanglement entropy as a function of $\delta/t_2$. For small values of  $\delta/t_2$, the entanglement entropy is zero. At $\delta/t_2=0.46$, it abruptly jumps to $\ln(4)$, corroborating the existence of a topological phase transition from a trivial to a topological phase. The value of $\ln(4)$ is expected because between $\delta/t_2=0.46$ and $\delta/t_2=1.49$, the system is in the topological (PH protected and PH broken) phase. At $\delta/t_2=1.49$, the entropy abruptly jumps again to zero, signalling a phase transition to a trivial phase. For $\delta/t_2>1.49$, the entropy increase logarithmically due to a stronger correlation between the subsystems in the PH-symmetry broken phase.   
	\begin{figure}[tbh]
		\centering
		\includegraphics[scale=0.6]{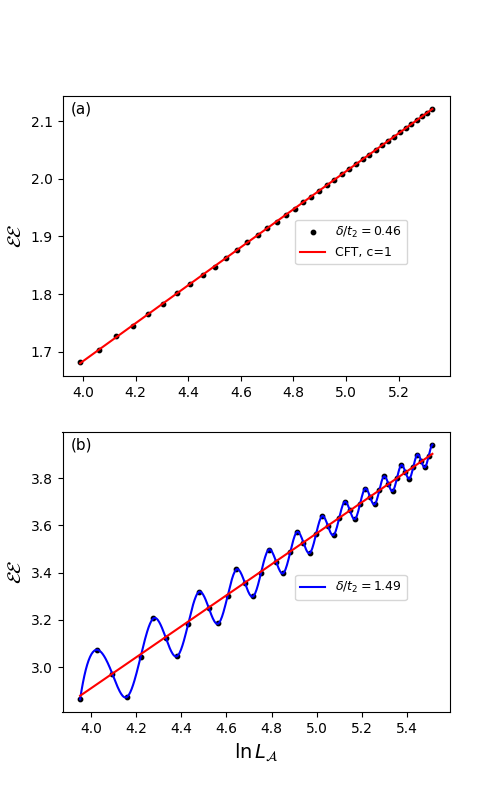}
		%	\captionsetup{justification=raggedright}
		\caption{Entanglement entropy of the nH-SSH model in terms of the logarithm of the system size $50<L<250$. (a) The black dots correspond to the entanglement entropy at $\delta/t_2=0.46$. The red line shows the linear behavior of the entanglement entropy in terms of $\ln{L_{\mathcal{A}}}$ with a slope of $1/3$, which corresponds to a central charge of $c=1$. (b) The black dots correspond to the entanglement entropy at  $\delta/t_2=1.49$. The blue line is a guide to the eye to better reveal the oscillations. The red line has a slope of 0.66 which corresponds to a central charge of $c=2$.}
		\label{Ent_Entropy_NHSSH}
	\end{figure}
	In Fig.~\ref{Ent_Entropy_NHSSH}(a), we show the entanglement entropy in terms of $\ln (L_{\mathcal{A}})$ at $\delta/t_2=0.46$. A logarithmic scaling of the entanglement entropy $\mathcal{EE} \sim (c/3)\ln(L_{\mathcal{A}})$ is observed for a subsystem size $L_{\mathcal{A}}$ with a central charge $c=1$, which is in agreement with the CFT description \cite{Pasquale-JSM2004,Cardy-NPB1988}. In Fig.~\ref{Ent_Entropy_NHSSH}(b), the entanglement entropy is plotted in terms of $\ln (L_{\mathcal{A}})$ at $\delta/t_2=1.49$. Oscillations are visible, although the amplitude and frequency decrease upon increasing the system size. In the thermodynamic limit, the oscillations will vanish. A linear behavior of the entanglement entropy as a function of the system size arise, which corresponds to a central charge $c=2$. In the literature, a CFT was found for a Parity-Time-symmetry broken phase in the gain and loss nH-SSH model. The central charge is negative $c=-2$ and the logarithmic scaling relation of the entanglement entropy with respect to the system size is $S_\mathcal{A}=(c/3)\ln \left[ \sin(\pi L_\mathcal{A}/L)\right]+\textnormal{const}$ \cite{Chang-PRR2020}.
	
	\subsection{Thermal entropy and heat capacity}\label{NHSSHES}
	Now, we investigate the thermal entropy, Eq.(\ref{3-6}) and the heat capacity, Eq.(\ref{3-7}), of the nH-SSH model at finite temperature. We calculate both quantities for three interesting values of the $\delta$ parameter: in the surroundings of the phase transition between the PH-symmetry protected trivial and the topological phase ($\delta/t_2=0.46$), at the non Bloch band collapse (PH-symmetry protected topological and PH-symmetry broken topological phases) $\delta/t_2= 1.1$, and at the phase transition between the PH-symmetry broken topological and trivial phases $\delta/t_2=1.49$. We calculate the entropy of the bulk and edge using Hill thermodynamics. The grand potential for the edge was obtained using the surrogate and the OBC Hamiltonian with $S=-\partial_T\Omega_{Edge}=-\partial_T(\Omega_{OBC}-\Omega_{sur})$. 
	\begin{figure}[tbh]
		\centering
		\includegraphics[scale=0.6]{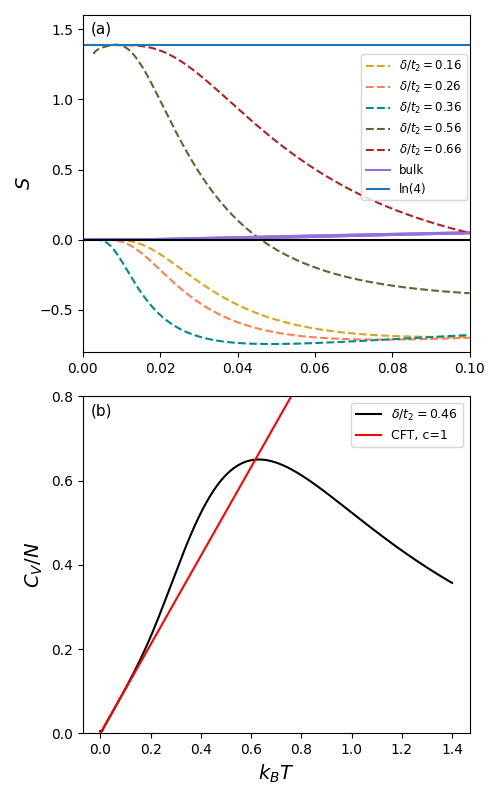}
		%	\captionsetup{justification=raggedright}
		\caption{Thermodynamic behavior of the nH-SSH model in the neighborhood of the phase transition between the PH-symmetry protected trivial and topological phases, at $\delta/t_2=0.46$, for a system size $L=200$. (a) Thermal entropy. The purple lines correspond to the bulk and the color dashed lines correspond to the edge, for different values of $\delta/t_2=\{0.16,0.26,0.36,0.56,0.66\}$. The blue line signals $\ln(4)$ for reference. (b) Specific heat of the bulk (black line) at $\delta/t_2=0.46$. The red line shows the linear behavior of the heat capacity at low temperatures, with a slope $C_v/Nk_B T \approx \pi/3$, which corresponds to a central charge of $c=1$.}
		\label{Entropy_NHSSHd1}
	\end{figure}
	\subsubsection{$\delta=0.46$}
	At the critical point $\delta/t_2=0.46$, the system undergoes a phase transition from a trivial to a topological phase. In both phases, PH-symmetry is preserved. The entropy for several values of delta in the neighbourhood of this critical point are plotted in Fig.~\ref{Entropy_NHSSHd1}(a). Although the bulk entropy always starts at zero and grows smoothly, the edge entropy changes significantly and allows us to detect the topological phase transition at $\delta/t_2 = 0.46$. For $\delta/t_2 < 0.46$, the entropy of the edge starts at zero and becomes negative upon increasing the temperature, whereas for $\delta/t_2 > 0.46$ it starts at $\ln 4$ for $T = 0$ and decays to zero as the temperature increases. One observes that the closer the value of $\delta$ is to the critical point $\delta_c/t_2 = 0.46$, the earlier the entropy falls off as one increases the temperature, see Fig.~\ref{Entropy_NHSSHd1}(a). At the critical point, the bulk heat capacity exhibits a linear behavior at low temperatures, in agreement with a CFT description with central charge $c = 1$. The entire behaviour is similar to the one observed for the Hermitian SSH model in Fig. 2 because the non-Hermitian SSH model is adiabatically connected to the Hermitian one around $\delta = 0$. The difference is that here the thermodynamic variables are plotted as a function of $\delta$ for a fixed value of $t_1/t_2$, whereas in 
Fig.~\ref{EntropyCv_SSH} the parameter driving the phase transition was $t_1/t_2$ because for the Hermitian SSH model $\delta = 0$. 	
	%\begin{figure}[H]
	%	\centering
	%	\includegraphics[scale=0.3]{HCvsTNHSSH400d0.46_p.png}
	%%	\captionsetup{justification=raggedright}
	%	\caption{Heat capacity of non-Hermitian SSH model (blue line) at phase transition between Hermitian trivial and Hermitian topological phases $\delta=0.46$. Red line shows a linear behavior of heat capacity for low $T$ temperatures with a slope $C_v/NK_B T \approx \pi/3$ which means a central charge of $c=1$}
	%	\label{Heat_NHSSHd1}
	%\end{figure}	
	\begin{figure}[bth]
		\centering
		\includegraphics[scale=0.6]{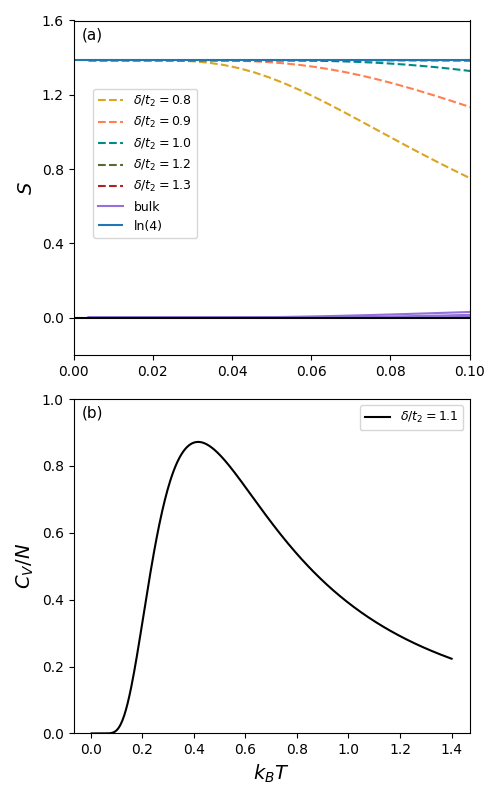}
		%	\captionsetup{justification=raggedright}
		\caption{Thermodynamic behavior of the nH-SSH model in the neighborhood of the phase transition between the topological phases with PH-symmetry protected and broken $\delta/t_2=1.1$. (a) Thermal Entropy. The purple lines correspond to the bulk, the color dashed lines correspond to the edge for different values of $\delta/t_2=\{0.8,0.9,1.0,1.2,1.3\}$. The blue line corresponds to $\ln(4)$. Edge and bulk contributions were computed for a system size $L=100$. (b) Specific heat of the bulk (black line) for a system size $L=200$ at the nBBC $\delta/t_2=1.1$.}
		\label{Entropy_NHSSHdnbbc}
	\end{figure}
	\subsubsection{$\delta=1.1$ (non-Bloch band collapse)}
	In Fig.~\ref{Entropy_NHSSHdnbbc}, we show the entropy and heat capacity plots in the neighbourhood of the nBBC point $\delta/t_2=1.1$. For $\delta/t_2 < 1.1$, the system is in the PH-symmetric topological phase. Hence, the edge entropy starts at $\ln 4$ and decays upon increasing the temperature, as observed for large values of $\delta$ in Fig.~\ref{Entropy_NHSSHd1}. However, for $\delta/t_2 > 1.1$, the edge entropy starts at $\ln 4$ and remains constant upon increasing the temperature with a slight increase, see Fig.~\ref{Entropy_NHSSHdnbbc}(a). Since this transition is between two different topological phases, the entropy always remains at $\ln 4$ for $T = 0$, but decreases in the PH-symmetric phase and remains constant (slightly increases) in the PH-symmetry broken phase at finite temperatures. The bulk entropy starts at zero and grows smoothly, as before. However, the heat capacity no longer exhibits a linear scale at low temperatures, but remains zero until a critical temperature, and grows afterwards, see Fig.~\ref{Entropy_NHSSHdnbbc}(b).  
	%\begin{figure}[H]
	%	\centering
	%	\includegraphics[scale=0.3]{HCvsTNHSSH400d1.1_p.png}
	%%	\captionsetup{justification=raggedright}
	%	\caption{Heat capacity of non-Hermitian SSH model (blue line) at non Bloch band collapse $\delta=1.1$.}
	%	\label{Heat_NHSSHdnbbc}
	%\end{figure}
%	
	\begin{figure}[tbh]
		\centering
		\includegraphics[scale=0.6]{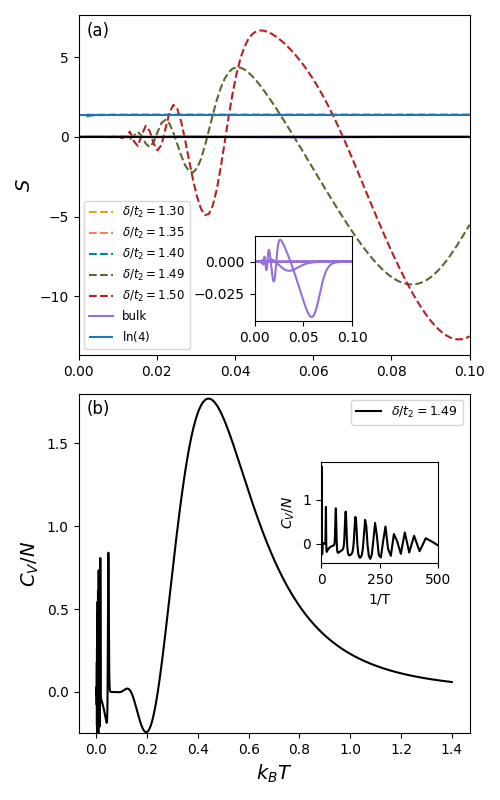}
		%	\captionsetup{justification=raggedright}
		\caption{Thermodynamic behavior of the nH-SSH model in the neighborhood of the phase transition between the PH-symmetry broken topological and trivial phase $\delta/t_2=1.49$. (a) Thermal Entropy. The purple lines correspond to the bulk, the color dashed lines correspond to the edge for different values of $\delta/t_2=\{1.19,1.29,1.39,1.59,1.69\}$. The blue line corresponds to $\ln(4)$. Inset shows oscillations of the bulk entropy at the critical point $\delta/t_2=1.49$. Edge contributions were computed with a system size $L=100$; bulk contributions were computed with a system size $L=200$ (b) Specific heat of the non-Hermitian SSH model for a system size $L=200$ (black line) at $\delta/t_2=1.49$. In the inset, the behavior of $C_v$ vs $1/T$ reveals periodic oscillations, characterizing an imaginary time crystal.} 
		\label{Entropy_NHSSHd2}
	\end{figure}
	\subsubsection{$\delta=1.49$}
	At the critical point $\delta/t_2=1.49$, the system undergoes a phase transition from a topological to a trivial phase. PH-symmetry remains broken in both phases. For $\delta/t_2 < 1.49$, the entropy starts at $\ln (4)$ and remains constant upon increasing the temperature, as observed before in Fig. \ref{Entropy_NHSSHdnbbc}. At $\delta/t_2 = 1.49$, there is a phase transition to the trivial phase, and the zero-temperature entropy of the edge jumps to zero. Now, upon increasing the temperature the bulk and edge entropy display an oscillatory behavior [see Fig.~\ref{Entropy_NHSSHd2}(a) and its inset] Interestingly, these oscillations are also visible in the bulk heat capacity, see Fig.~\ref{Entropy_NHSSHd2}(b), where the inset reveals that the oscillations are periodic in $1/T$. Therefore, they characterize the emergence of an imaginary time crystal \cite{WilczekTC, AroucaMarino-QF2022}. This phase appears when the resonance condition
	\begin{equation}
		i \hbar\omega_n=E_{\text{surr}, \pm}(k),
	\end{equation}
	with $\omega_n=(2n+1) \pi k_B T/\hbar$ the Matsubara frequencies for a quantum fermionic system, is satisfied for any of the modes. 
	Since the Matsubara frequencies $\omega_n$ are real, this condition is just obeyed when there are modes with purely imaginary energies. From the expression of the surrogate Hamiltonian, Eq.~\eqref{eq_E_surr}, we see that this is obtained when $k=\pm \pi/2$, \footnote{Notice that the term multiplying $\cos(k)$ inside the square root is always imaginary, what makes that the energies with $k\neq \pm \pi/2$ have a finite real part.} $$1+t_1^2-\delta^2< 0 \rightarrow \delta>\sqrt{1+t_1^2},$$ which implies that the system is in the PH broken trivial phase. Moreover, there will be peaks in the thermodynamic potentials for temperatures that satisfy
	\begin{equation}
		(2n+1) \pi k_B T=\sqrt{\delta^2-t_1^2-1},
	\end{equation}
	such that the period of the peaks in inverse of temperature is given by $2\pi k_B/\sqrt{\delta^2-t_1^2-1}$. The oscillations in temperature are related to the spontaneous breaking of translation symmetry in imaginary time and characterize the emergence of an imaginary time crystal, a phase conjectured by Wilczek \cite{WilczekTC} and shown to be present in non-Hermitian systems with purely imaginary modes in Ref.~\cite{AroucaMarino-QF2022}. Here, we show that they also arise in the realm of the non-reciprocal SSH model, within the trivial PH-symmetry broken phase. Notice that although the peaks have a periodicity, the specific heat decreases with 1/T.

	\section{Conclusions}\label{conclusions}
	
	Using the Hill thermodynamics formalism, we identified topological phase transitions for the SSH and nH-SSH models at finite temperature, by investigating the thermal entropy and heat capacity for the bulk and edge. In addition, the entanglement entropy of the nH-SSH model was studied to identify topological phases at zero temperature.  
	
	We found that the thermodynamics of the nH-SSH model is in agreement with the results of the entanglement entropy for all phases. At the critical point of the transition between the PH-symmetry protected trivial and topological phases, both, the entanglement entropy as a function of the subsystem size and the the heat capacity at low temperature show that the model is a CFT fermionic system. We also showed that the entanglement entropy and the thermal entropy of the edge host fermionic zero modes in the PH-symmetry protected topological phase.
	
	At the nBBC critical point, i.e., the transition between the PH-symmetry protected and broken topological phases, we found that the non-Hermitian SSH model is not a CFT. As the system is always in a topological phase, the entanglement entropy and the thermal entropy remain at $\ln(4)$, showing that these phases host fermionic zero modes, independently of the PH-symmetry being broken or preserved.
	
	Finally, we investigated the relations of the entanglement entropy as a function of the subsystem size and the heat capacity as a function of the inverse of the temperature at the critical point where both, the topological and trivial phases have a broken PH-symmetry. It is interesting that oscillations were observed for both, the entanglement entropy and heat capacity. For the latter this means that there is the emergence of an imaginary time crystal, whereas for the entanglement entropy, it was not possible to identify a clear periodicity.
	
	Although we obtain interesting results while investigating an effective non-Hermitian model, it is possible that more realistic descriptions of quantum open systems, like the ones based on the Lindblad master equation or on the Caldeira-Legget formalism, could exhibit qualitatively different features. Therefore, a natural extension of our work would be to investigate the fate of these transitions at finite temperatures using these formalisms. Furthermore, the physical interpretation of the non-integer order of the phase transition obtained for the transition between the PH-broken phases is an interesting venue of research.

	\section{Acknowledgments}
	
	The authors thanks to D.S. Quevedo and C. Ortega-Taberner for the valuable help implementing the numerical methods. The authors thanks also A. Moustaj, S.N. Kempkes, A. Quelle, L. R{\o}dland, P. Calabrese, and G. Mussardo for valuable discussions. DFM-A acknowledges funding from the Colombian Ministry of Science (Minciencias) and from the Delta Institute for Theoretical Physics (DITP) consortium, a program of the Netherlands organization for Scientific Research (NWO) that is funded by the Dutch Ministry of Education, Culture and Science. RA is thankful to the Knut and Alice Wallenberg foundation for funding.  
	
	\bibliography{Bibliography_prb}% Produces the bibliography via BibTeX.
\end{document}